\documentclass[
12pt,
3p,
review,
]{elsarticle}
\usepackage{hyperref}
\usepackage{graphicx,graphics}
\usepackage{placeins}
\usepackage{amsmath}
\usepackage{amssymb}
\usepackage{mathtools}

\usepackage{array}

\usepackage{hyperref}
\usepackage{color}
\usepackage{soul}
\usepackage{dcolumn}% Align table columns on decimal point
\usepackage{bm}% bold math
\usepackage{todonotes}
\usepackage[utf8]{inputenc}
\usepackage[english]{babel}
\usepackage{libertine}
\usepackage[caption=false]{subfig}
\usepackage{pgfplots}
\usepackage{booktabs}
\usepackage{upgreek}

\usepgfplotslibrary{polar}
\pgfplotsset{compat=1.9}
\usetikzlibrary{calc}

\usepackage{layouts}

\addto\captionsenglish{}

\newcommand{\vphi}{\mbox{{\bm{$\phi$}}}}

\newcommand{\vv}[1]{\boldsymbol{#1}}
\newcommand{\n}{\vv{\nabla}}

\newcommand{\diff} [1]{\mathrm{d}{#1}} 
\newcommand{\phia}{\phi_{\alpha}}
\newcommand{\phib}{\phi_{\beta}}
\newcommand{\gab}{\gamma_{\alpha \beta}}

\newcommand{\qab}{\vv{q}_{\alpha \beta}}
\newcommand{\C}{\vv{\mathcal{C}}}

\let\originaleps=\epsilon
\let\epsilon=\varepsilon
\let\varepsilon=\originaleps
\usepackage{stmaryrd}
\newcommand{\ljump}{\llbracket}
\newcommand{\rjump}{\rrbracket}

\newcommand{\pace}{{\textsc{Pace3D}}}

\definecolor{mydark_blue}{RGB}{0, 0, 139}
\definecolor{myblue}{RGB}{0, 0, 255}
\definecolor{mycyan}{RGB}{0, 255, 255}  
\definecolor{mygreen}{RGB}{0, 255, 0}
\definecolor{myyellow}{RGB}{255, 255, 0}
\definecolor{myred}{RGB}{255, 0, 0}
\definecolor{mydark_red}{RGB}{139, 0, 0}
\definecolor{myblack}{RGB}{0, 0, 0}

\definecolor{BRY_1}{RGB}{  0,  0,255}
\definecolor{BRY_2}{RGB}{127,  0,127}
\definecolor{BRY_3}{RGB}{255,  0,  0}
\definecolor{BRY_4}{RGB}{255,127,  0}
\definecolor{BRY_5}{RGB}{255,255, 85}

\newcommand{\TT}{\vv{\mathcal{T}}}

\journal{Materialia}

\begin{document}
Postprint \glqq Influence of stress-free transformation strain on the autocatalytic growth of bainite: A multiphase-field analysis\grqq{} by E. Schoof, P.G. Kubendran Amos, D. Schneider, B. Nestler, accepted for publication in \textit{Materialia} on Feb 04, 2020. \\
This work is licensed under CC-BY-NC-ND. \\
Original publication: https://doi.org/10.1016/j.mtla.2020.100620 

\newpage

\begin{frontmatter}
\title{Influence of stress-free transformation strain on the autocatalytic growth of bainite: A multiphase-field analysis}

\author[mymainaddress]{Ephraim Schoof\corref{mycorrespondingauthor}\fnref{fn1}}
\cortext[mycorrespondingauthor]{Ephraim Schoof}
\ead{ephraim.schoof@kit.edu}

\author[mymainaddress]{P.G. Kubendran Amos\fnref{fn1}}
\author[mymainaddress,mysecondaryaddress]{Daniel Schneider}
\author[mymainaddress,mysecondaryaddress]{Britta Nestler}

\fntext[fn1]{The authors contributed equally.}

\address[mymainaddress]{Institute for Applied Materials (IAM-CMS), Karlsruhe Institute of Technology (KIT),\\
Strasse am Forum 7, 76131 Karlsruhe, Germany}
\address[mysecondaryaddress]{Institute for Digital Materials Science (IDM), Karlsruhe University of Applied Sciences,\\
Moltkestr. 30, 76133 Karlsruhe, Germany}

\begin{abstract} 
Analytical treatments, formulated to predict the rate of the bainite transformation, define autocatalysis as the growth of the subunits at the bainite-austenite interface.
Furthermore, the role of the stress-free transformation strain is often translated to a thermodynamic criterion that needs to be fulfilled for the growth of the subunits.
In the present work, an elastic phase-field model, which elegantly recovers the sharp-interface relations, is employed to comprehensively explicate the effect of the elastic energy on the evolution of the subunits.
The primary finding of the current analysis is that the role of eigenstrains in the bainite transformation is apparently complicated to be directly quantified as the thermodynamic constraint.
It is realized that the inhomogeneous stress state, induced by the growth of the primary subunit, renders a spatially dependent ill- and well-favored condition for the growth of the secondary subunits.
A favorability contour, which encloses the sections that facilitate the elastically preferred growth, is postulated based on the elastic interaction.
Through the numerical analyses, the enhanced growth of the subunits within the favorability-contour is verified.
Current investigations show that the morphology and size of the elastically preferred region respectively changes and increases with the progressive growth of the subunits.

\end{abstract}


\end{frontmatter}

% \linenumbers

\section{Introduction}
Despite being widely used and extensively investigated, the phase transformations associated with the iron-carbon system are far from reaching an absolute consensus.
Particularly, the debate pertaining to the bainite transformation is wide and active, with each school holding a position that marginally overlaps~\cite{hehemann1972debate}.
In other words, while some regard the decomposition of austenite to bainite as a reconstructive transformation, primarily governed by diffusion~\cite{purdy1984overview,aaronson1990bainite,van2002discussion}, the contradicting view postulates it as a displacive transformation, wherein the para-equilibrium is established between the phases by the exclusive migration of carbon~\cite{hehemann1970phase,bhadeshia1981rationalisation,rees1992bainite}.
The significant differences between the perceived understanding of the transformation and the apparent consent in some aspects have extensively been elucidated elsewhere~\cite{fielding2013bainite}.
Although seemingly valid arguments are rendered by these opposing views, owing to their success in quantitative predictions, which have been verified experimentally, the displacive position is relatively preferred for formulating the kinetics of the bainite transformation.

According to the displacive theory, the growth of the bainite sheave is predominantly diffusionless~\cite{bhadeshia1982bainite}.
However, unlike the martensite transformation, wherein a given variant progressively grows until it reaches a barrier like a grain boundary, the bainite transformation reflects the cumulative growth of individual subunits.
These numerous subunits, which constitute the bainite sheave, nucleate and grow akin to the martensite lath.
The diffusionless growth of the individual subunits is impeded by the plastic accommodation at the tip~\cite{bhadeshia1979bainite}.
This characteristic growth renders a definite shape and aspect ratio to the subunits.
Despite the curbed growth of the subunits, the increased rate of transformation in bainite is achieved by the complementing \lq autocatalytic\rq \thinspace nucleations.

\subsection{Generalized framework for the estimation of the transformation kinetics}\label{Sec:framework}

To understand the role of the autocatalytic nucleation in the transformation kinetics of bainite, and subsequently analyze the influence of strain on this mode of nucleation, which is the primary motive of the present study, it is vital to describe the theoretical framework adopted to estimate the transformation rate.
To that end, and to motivate the current analysis, different models developed over a period of thirty years are concisely, yet coherently, discussed in this section.

Initial reports on the kinetics of the bainite transformation adopt the Johnson-Mehl-Avrami-Kolmogorov~(JMAK) relation to formulate the temporal change in the volume fraction of bainite~\cite{bhadeshia1982bainite,chester1997mathematical}.
It is argued that the JMAK framework, particularly the introduction of the \textit{extended volume}, appropriately negates the contribution of the \lq phantom nuclei\rq~in the decomposed austenite~\cite{rees1992bainite}.
However, recent investigations have shown that the outcomes of regular formulations, devoid of any extended volume consideration, can be consistent with the experimental observations~\cite{van2008modeling,ravi2016exploring}. 
Irrespective of the framework, JMAK or otherwise, two primary aspects of the formulation remain unchanged.

The first one is the proportionality between the temporal change in the volume fraction of bainite and the nucleation rate of the subunits.
Owing to the nature of the transformation, the subunits, which are finite and significantly smaller than the bainite sheave, achieve their definite size at an exceedingly faster rate.
Resolving this growth period, which is infinitesimal when compared to the total nucleation, is redundant and practically an arduous task. 
Therefore, the growth kinetics of bainite is expressed in relation to the nucleation rate of the subunits.

The other primary aspect, which is independent of the framework, is the activation energy associated with the formulation of the transformation kinetics. 
Based on the influence of the driving force on the bainite start temperature,~$B_\text{s}$, a linear dependency between the activation energy and the driving force is assumed~\cite{bhadeshia1981rationalisation}.
Correspondingly, the thermodynamic driving force is introduced in the formulation through the activation energy.

Overlooking the contribution of the driving force, without losing the generality, the framework for ascertaining the transformation kinetics is written as
\begin{align}\label{eq:frame1a}
 \frac{\partial v^{\alpha'}}{\partial t}\propto u_{\text{sb}}I,
\end{align}
where~$v^{\alpha'}$ and~$u_{\text{sb}}$ correspond to the volume fraction of bainite and to the volume of a subunit, respectively~\cite{bhadeshia1982bainite,rees1992bainite}. 
The nucleation rate per unit volume is expressed with~$I$.
This formulation assumes that the subunit attains the volume~$u$ instantly after the nucleation.
The nucleation rate~$I$ in Eqn.~\eqref{eq:frame1a} includes the primary nucleation associated with the preexisting austenite grain boundaries and the autocatalytic nucleation.
Correspondingly, the overall nucleation rate is expressed as
\begin{align}\label{eq:frame1b}
I(S_{\text{GB}},c)=I_{0}(S_{\text{GB}})[1+{\mathcal{B}}(c)v^{\alpha'}],
\end{align}
where~$I_{0}$ is the primary nucleation rate which is dependent on the austenite-austenite grain boundary area~$S_{\text{GB}}$.
The autocatalytic nucleation is introduced through the dimensionless factor~$\mathcal{B}$.
Since the autocatalytic nucleation is conventionally defined as the nucleations restricted to the austenite-bainite interfaces, the bainite volume fraction is included in Eqn.~\eqref{eq:frame1b}.
Moreover, in formulations wherein similar activation energies are considered for primary and autocatalytic nucleation~\cite{rees1992bainite}, the influence of concentration is introduced as follows through the factor
$\mathcal{B}$:
\begin{align}\label{eq:frame1c}
 {\mathcal{B}}(c)=\lambda_1(1-\lambda_2 c).
\end{align}
Here,~$c$ is the mole fraction of carbon, while~$\lambda_1$ and~$\lambda_2$ are empirical constants.
The subsequent extension of this approach involves introducing the number of nucleation sites~\cite{singh1998phase,singh2012mechanisms}.

The nucleation rate, which dictates the rate of the bainite transformation, is related to the corresponding activation energy through a set of empirical constants.
These constants are determined by fitting the predictions of the formulation to the experimental results.
In order to replace the empirical constants, the approach is reformulated, and the primary nucleation rate is expressed as
\begin{align}\label{eq:frame2a}
 I_{0} \propto N_{0}^{\alpha'}(S_{\text{GB}},u_{\text{SU}})\mathcal{V},
\end{align}
where~$N_{0}^{\alpha'}$ is the number density of the nucleation sites and~$\mathcal{V}$ is the frequency of attempts to form stable nuclei.
In Eqn.~\eqref{eq:frame2a}, the density of the potential sites is expressed as a function of the grain boundary area~($S_{\text{GB}}$) and the thickness of the subunit~($u_{\text{SU}}$).
Despite the introduction of a thermodynamically pertinent parameter, i.e., nucleation density and attempt frequency, the overall nucleation rate is still written as 
\begin{align}\label{eq:frame2b}
 I \propto \underbrace{ N_{0}^{\alpha'}(S_{\text{GB}},u_{\text{SU}}) + \mathcal{B}(c)\chi(I_{0})}_{N_{T}^{\alpha'}},
\end{align}
wherein the autocatalytic nucleation is introduced through the nondimensional factor~$\mathcal{B}(c)$.
The parameter~$\chi(I_{0})$ in Eqn.~\eqref{eq:frame2b} includes the activation energy, which is similar for both primary and autocatalytic nucleation.
Furthermore, in consistence with the framework, the overall nucleation rate is expressed in relation to the total density of the nucleation site~$N_{T}^{\alpha'}$~\cite{singh1998phase,opdenacker2001rate}. 

Theoretical treatments, wherein the nucleation rates are formulated analogous to Eqn.~\eqref{eq:frame1b} (or Eqn.~\eqref{eq:frame2b}), inherently assume a generalized activation energy for all nucleations, irrespective of the location of the site (grain boundary or~$\alpha'\gamma$-interface).

Therefore, an alternate approach followed the conventional framework, wherein the overall nucleation rate is expressed as
\begin{align}\label{eq:frame3a}
 I=I_{0}(N_{0}^{\alpha'},S_{\text{GB}},Q_{\text{GB}})+\tilde{\mathcal{B}}I_{\text{A}}(I_{0},N_{A}^{\alpha'},Q_{\text{A}}),
\end{align}
where the primary nucleation rate $I_{0}$ is distinguished from the autocatalysis $I_{\text{A}}$~\cite{gaude2006new}.
Although a dimensionless prefactor~$\tilde{\mathcal{B}}$ is involved in Eqn.~\eqref{eq:frame3a}, different activation energies, $Q_{\text{GB}}$ and~$Q_{\text{A}}$, are correspondingly adopted for grain boundary and~$\alpha'\gamma$-interface nucleation. 
Moreover, the interaction between the different activation energies, particularly during the autocatalytic nucleation, is implicitly considered in this formulation~\cite{gaude2006new}.
In Eqn.~\eqref{eq:frame3a}, the potential nucleation sites for autocatalysis are included as~$N_{A}^{\alpha'}(\eta_{A},\eta_{B})$, where~$\eta_{A}$ and~$\eta_{B}$ are the dimensions of the bainite subunits.
In this framework, although the influence of the invariant plane-strain associated with the bainite transformation on the autocatalytic nucleation is realized, a lack of sufficient elucidation of this interaction is conceded.

Recent works have seemingly adopted and extended the delineation of the overall nucleation in Eqn.~\eqref{eq:frame3a}, by coherently removing the dimensionless constant~$\tilde{\mathcal{B}}$ and explicitly treating the interactions between the different activation energies $\Delta Q^{*}=Q_{\text{GB}}-Q_{\text{A}}$~\cite{van2008modeling,ravi2017bainite}.
Moreover, the effect of the grain and the subunit size on the nucleation rate is introduced through the number of potential sites in the extended formulations~\cite{van2012physically,ravi2016exploring}.
Even though these advanced treatments appropriately relate the interactions between the activation energies $\Delta Q^{*}$ to the autocatalytic nucleation, the lack of a definite contribution of the strain to the transformation rate is recognized.

The brief review of the existing models indicates that, owing to a general absence of a convincing understanding of the influence of the stress-free transformation strain, both primary and autocatalytic nucleation rates are predominantly represented as a function of the potential nucleation site, the geometry and the size of the grains and subunits.
In this study, a multiphase-field model is therefore adopted to explain the effect of the elastic interaction on the nucleation and growth of bainite subunits. 

\subsection{Other frameworks}

In the previous Sec.~\ref{Sec:framework}, an overview of different analytical models, which adhere to a generalized framework characterized by Eqn.~\eqref{eq:frame1a}, is elucidated to explain the common inadequacy pertaining to the role of the elastic strain.
Several approaches exist which provide a convincing delineation of the transformation kinetics, while deviating from the generalized treatment~\cite{tszeng2000autocatalysis,matsuda2004kinetics}.
Most of these techniques are directed towards reducing the number of empirical constants associated with the formulation, which are ascertained by comparing the predictions with the experimental results.
Despite the differences in the framework, these works also realize the significant role of the characteristic stress-free transformation strain and concede its absence. 

Although not explicitly stated, a pivotal effect of the strain, which favors the autocatalytic nucleation at the tip of the subunit, is acknowledged and even implemented in some of these works~\cite{tszeng2000autocatalysis,gaude2006new}.
In the present analysis, these energetically favored spots for the autocatalytic events are quantitatively described.

\section{Simulation setup}

Phase-field models are increasingly used to analyze a wide range of microstructural evolutions, including solidification~\cite{chen2002phase,steinbach2009phase,nestler2011phase}, solid-state phase transformations~\cite{mushongera2018phase,amos2018mechanisms} and curvature-driven morphological evolutions~\cite{amos2018phase2,amos2018volume,amos2018phase}.
Generally, an evolution in a theoretical framework is comprehended by monitoring the migration of the interface.
Therefore, with the intricacy of the evolving structures, the corresponding formulation becomes convoluted.
The phase-field technique averts this complexity by introducing a scalar variable, which assumes a constant value with a given phase and smoothly varies across the interface.
Correspondingly, the evolution then reflects the spatio-temporal changes in the scalar variable, called phase field.
An absolute compliance with the physical laws and conventional sharp-interface solutions, despite the introduction of a diffuse interface by the spatially varying scalar variable, is ensured either by asymptotic analysis~\cite{wheeler1992phase,wang2008asymptotic,amos2019understanding} or by appropriately treating the thermodynamic functions across the interface~\cite{tiaden1998multiphase,kim1998interfacial,kim1999phase}. 
Moreover, this ability of the phase-field approach to quantitatively encompass the underlying physics, while being versatile, is a reason for its growing recognition.

Through the incorporation of the elastic driving force, phase-field models have been employed to simulate phase changes like the martensite transformation~\cite{yeddu2012three,heo2014phase,schoof2018multiphase,schoof2018multiphaseplast} and the growth of Widmanst\"{a}tten ferrite~\cite{cottura2014phase,amos2018chemo}.
In these works, the microstructural evolution, dictated by the elastically favored autocatalytic nucleation, and the growth of different variants is elegantly discussed based on the phase-field simulations.
Moreover, by including appropriate chemical driving forces, this approach has been extended to quantitatively predict the variant selection during the~$\alpha-$precipitation in titanium alloys~\cite{shi2013variant,qiu2016effect}.
However, the autocatalysis associated with the bainite transformation significantly differs from the elastically preferred variant growth.
Particularly, it is realized that the subunits, which nucleate and grow autocatalytically in a bainite sheave, apparently exhibit a similar orientation relation~\cite{bhadeshia2019bainite}.
This characteristic orientation relation indicates that the subunits within a bainite sheave pertain to a single variant. 
Despite the several analytical and numerical modelings of bainite transformation, e.g.~\cite{arif2013phase,hueter2015multiscale,dusing2016simulation,song2018multiphase,dusing2018coupled}, an extensive analysis of the elastically governed autocatalytic evolution of a single variant is largely impending.
Given the influence of the autocatalysis in the formulation of the transformation kinetics, as described in Sec.~\ref{Sec:framework}, the influence of the stress-free transformation strain on the nucleation and growth of a subunit is investigated by adopting a thermodynamically consistent, multiphase-field model coupled with linear elasticity.
A brief and contextual description of this multiphase-field model is given in this section.
For a detailed description, the readers are referred to Refs.~\cite{steinbach1996phase,schneider2015phase,schneider2018small}.

\subsection{Multiphase-field model}

\subsection{Overall energy density}

Based on the spatial nature of the phase field, the entire system can be separated into bulk phases and interface regions.
In other words, the regions wherein the scalar variables assume a constant value are treated as the bulk phases, while the interfaces are the diffuse regions characterized by the spatial change in the phase field, separating the bulk phases.
Correspondingly, the overall energy density of the system is written in the form of a Ginzburg-Landau functional:
\begin{align}\label{eq:functional}
  \mathcal{F}(\vphi, \n \vphi,\vv{S}) &= \mathcal{F}_\text{intf}(\vphi, \n \vphi) + \mathcal{F}_\text{bulk}(\vphi, \vv{S}).
\end{align}
Here, the contribution of the bulk phases and the interfaces is respectively included into~$\mathcal{F}_\text{intf}(\vphi, \n \vphi)$ and~$\mathcal{F}_\text{bulk}(\vphi, \vv{S})$.
Since the model is formulated for a multiphase setup, the phase field is considered as an $N$-tuple, $\vphi = \{\phia,\phib\dots\phi_{N}\}$, representing all~$N$-phases~\cite{steinbach1999generalized,nestler2005multicomponent}.
Moreover, in Eqn.~\eqref{eq:functional}, $\vv{S}$ collectively denotes the fundamental variables which dictate the bulk-phase contribution.

Using the individual free energy densities, the functional describing the energy state for the present system of volume~$V$ can be defined as
\begin{align}\label{eq:overall_freeEnergy}
\begin{split}
 \mathcal{F}(\vphi, \n \vphi,\vv{\varepsilon},c,T) &= \int_{V}\Big[ W_{\text{intf}}(\vphi, \n \vphi) + W_{\text{bulk}}(\vv{S}, \vphi) \Big]\diff V \\ 
 &=\int_{V}\Big[W_{\text{gr}}(\vphi, \n \vphi) + W_{\text{ob}}(\vphi)+ W_{\text{el}}(\vv{\varepsilon}, \vphi) + W_{\text{ch}}(c,T, \vphi)\Big]\diff V.
 \end{split}
\end{align}
Owing to the nature of the transformation analyzed in this work, the bulk free energy density ($W_{\text{bulk}}(\vv{S}, \vphi)$) is formulated to encompass elastic and chemical energy densities, $W_{\text{el}}(\vv{\varepsilon}, \vphi)$ and~$W_{\text{ch}}(c,T, \vphi)$, respectively~\cite{schoof2018multiphase}.
The generalized variable $\vv{S}$ in Eqn.~\eqref{eq:overall_freeEnergy} includes the constitutive variables~$\vv\varepsilon$, $c$ and~$T$, which correspond to the local strain, the mole fraction of carbon and the temperature.

\subsubsection{Interface contribution}

The interfacial energy contribution in Eqn.~\eqref{eq:overall_freeEnergy} comprises two components.
These are the gradient energy density $W_{\text{gr}}(\vphi, \n \vphi)$ and the potential energy density $W_{\text{ob}}(\vphi)$, which together form a diffuse interface with a defined width of the transition region.
In the current multiphase model, based on the fundamental works of Refs.~\cite{steinbach1999generalized,nestler2005multicomponent,steinbach2009phase}, the gradient energy density is written as
\begin{align}\label{eq:a}
 W_{\text{gr}}(\vphi, \n \vphi) = \epsilon a(\vphi, \n \vphi) = \epsilon \sum_{\alpha < \beta}^{N} \gab \underbrace{|{\phi_{\alpha} \bm{\nabla}{\phi}_{\beta}} - {\phi_{\beta} \bm{\nabla}{\phi}_{\alpha}}|^2}_{:=|\qab|^2},
\end{align}
where~$\epsilon$ is the length scale parameter, which determines the width of the diffuse interface, and~$\gab$ is the energy of the interface, separating phase~$\alpha$ and~$\beta$.
In Eqn.~\eqref{eq:a},~$\qab$ furthermore represents the generalized gradient vector~\cite{tschukin2017concepts}.

During the evolution, a constant value in the bulk phases is ensured by penalizing the phase field.
Generally, this is achieved by adopting a well-type function~\cite{warren1995prediction,suzuki2002phase}.
Given its numerical accuracy, the  obstacle-type potential in the present approach is involved in assigning definite values~\cite{oono1988study,perumal2018phase2}.
Moreover, the efficiency of the potential density $W_{\text{ob}}(\vphi)$ is enhanced by devising a Gibbs simplex of the form
\begin{equation}\label{eq:gibbs_simplex}
 \mathcal{G}=\left\{ \vv{\phi} \in \mathbb{R}^N: \sum_\alpha \phi_\alpha = 1, \phi_\alpha \geq 0 \right\}.
\end{equation}
By imposing the penalizing criterion through the Gibbs simplex, the potential density is formulated as 
\begin{equation}\label{eq:potential_energy}
 W_{\text{ob}}(\vphi)=\frac{1}{\epsilon}\omega(\vphi)= 
 \begin{cases}
 \frac{16}{\epsilon\pi^2} \underset{\alpha < \beta}{\sum} \gab \phia \phib + \frac{1}{\epsilon}\underset{\alpha < \beta < \delta}{\sum} \gamma_{\alpha \beta \delta} \phia \phib \phi_{\delta},&  ~\vphi \in \mathcal{G}\\
 \infty &  ~\vphi \notin \mathcal{G}.
 \end{cases}
\end{equation}
Accordingly, when the phase field moves out of the simplex~$\mathcal{G}$, the potential density sharply increases to~$\infty$.
This abrupt change in the potential density ensures that any deviation from the criterion~$\sum_\alpha \phi_\alpha = 1$ across the interface is extremely expensive.
Through the parameter~$\gamma_{\alpha \beta \delta}$, the third order term in Eqn.~\eqref{eq:potential_energy} furthermore averts the formation of spurious third phases~\cite{nestler2005multicomponent,perumal2017phase}.

\subsubsection{Bulk contribution}

In the current approach, as indicated in Eqn.~\eqref{eq:overall_freeEnergy}, the chemical and elastic energy densities constitute the bulk contribution.
The elastic free energy density~(Helmholtz) of the system is conventionally expressed as
\begin{align}\label{eq:elastic1}
 W_{\text{el}}(\vv{\varepsilon}, \vphi) & =\sum_{\alpha}W_{\text{el}}^{\alpha}(\vv{\varepsilon}^{\alpha})h(\phia) \\ \nonumber
 & = \sum_{\alpha} \frac{1}{2}\left[ (\varepsilon_{ij}^{\alpha}-\tilde{\varepsilon}_{ij}^{\alpha})\cdot \mathcal{C}_{ijkl}^\alpha (\varepsilon_{kl}^{\alpha}-\tilde{\varepsilon}_{kl}^{\alpha})  \right]h(\phia).
\end{align}
Here,~$W_{\text{el}}^{\alpha}(\vv{\varepsilon})$ is the energy density of an individual phase~$\alpha$ and~$h(\phia)$ is the corresponding interpolation function~\cite{khachaturyan1983theory}.
The overall elastic energy density of the system, $W_{\text{el}}(\vv{\varepsilon}, \vphi)$, is formulated by interpolating the elastic free energies of the separate phases.
Moreover, in Eqn.~\eqref{eq:elastic1}, $\varepsilon_{ij}^{\alpha}$, $\tilde{\varepsilon}_{ij}^{\alpha}$ and~$\mathcal{C}_{ijkl}^{\alpha}$ represent the total strain, the inelastic strains and the stiffness tensor, respectively.

The formulation in Eqn.~\eqref{eq:elastic1} often follows the assumption that either stress~\cite{spatschek2007phase,mennerich2011phase} or strain~\cite{steinbach2006multi,apel2009virtual} is constant across the interface.
Correspondingly, the continuous variable is identified, and the entire model is appropriately derived.
However, it has been clearly shown that in certain conditions, such a treatment of the variables contributes an excess energy to the interface, which unfavorably influences the overall kinetics of the evolution~\cite{schneider2018small}.
For instance, under uniaxial loading along the normal direction, the equal stress treatment in the transition region introduces excess energy into the interface.
The apparent failure to recover the physical interface energy ultimately compromises the quantitative nature of the results.
Following Refs.~\cite{schneider2015phase,schneider2018small}, the present approach therefore begins by identifying specific variables which do not contribute to any excess energy when treated as a continuous variable across the diffuse interface.

According to the Hadamard condition, the jump of the deformation gradient across a singular surface is expressed as the outer product of a vector and its normal. 
Written in terms of the total strain~$\vv\varepsilon$, this condition reads
\begin{align}\label{eq:strain_normal}
 \ljump \vv{\varepsilon} \rjump = \ljump \vv{\varepsilon}_{\text{el}} +  \vv{\tilde{\varepsilon}} \rjump=\frac{1}{2}(\vv{a} \otimes \vv{n}^{s} + \vv{n}^{s} \otimes \vv{a}),
\end{align}
where~$\vv{\varepsilon}_{\text{el}}$ and~$\vv{\tilde{\varepsilon}}$ correspond to the elastic and inelastic strain.
Furthermore, considering that the forces are balanced across the sharp interface of a singular surface, the jump in the stress along the normal direction is expressed as
\begin{align}\label{eq:force_balance1}
 \llbracket \vv{\sigma}\rrbracket_{\vv{n}} = \vv{0}.
\end{align}
Based on the characteristic jump conditions in Eqns.~\eqref{eq:strain_normal} and~\eqref{eq:force_balance1}, the normal components of the stresses are considered as continuous variables.

The delineation for identifying appropriate continuous variables explains the critical role of the normal vector.
In a multiphase setup, the normal vector should be ascertained in an efficient manner, so as to ensure the optimized use of the computational resources.
A scalar field 
\begin{align}\label{eq:n_aus_M}
M(\vv\phi)=\sum_{\alpha<\beta}\phia\phib
\end{align}
is defined in the present model to efficiently determine the normal vector.
The normal vector, based on this scalar field, is subsequently ascertained by 
\begin{align}\label{eq:n_aus_M1}
 \vv n(M(\vv\phi)) = \frac{\vv \nabla M(\vv\phi)}{|\vv \nabla M(\vv\phi)|}.
\end{align}
The efficiency of the technique of recognizing the normal vector is extensively discussed in Ref.~\cite{schneider2018small}.

Having identified the potential continuous variables, based on the jump conditions, the corresponding components of the stresses and strains are distinguished from the phase-dependent counterparts, so as to appropriately formulate the elastic free energy density.
To achieve an effective separation, a coordinate system of the base~$\vv{B}=\{\vv{n},\vv{t},\vv{s}\}$, consisting of a normal vector, $\vv{n}$, and two tangential vectors, $\vv{t}$ and~$\vv{s}$, is considered.

Based on Eqn.~\eqref{eq:strain_normal}, which results from the jump condition, the strain components in the Voigt notation are written as
\begin{align}\label{eq:strainB_voigt}
 \vv{\varepsilon}_{\vv{B}}^{\alpha} = \vv{\varepsilon}_{\text{el}:{\vv{B}}}^{\alpha} +  \vv{\tilde{\varepsilon}}_{\vv{B}}^{\alpha} = \left[ (\vv{\varepsilon}_{\text{el}:\vv{n}}^{\alpha}+  \vv{\tilde{\varepsilon}}_{\vv{n}}^{\alpha}), (\vv{\varepsilon}_{\text{el}:\vv{t}}+  \vv{\tilde{\varepsilon}}_{\vv{t}})   \right]^{T},
\end{align}
where the phase-dependent and continuous components are collectively and respectively represented by~$(\vv{\varepsilon}_{\text{el}:\vv{n}}^{\alpha}+  \vv{\tilde{\varepsilon}}_{\vv{n}}^{\alpha})$ and~$(\vv{\varepsilon}_{\text{el}:\vv{t}}+  \vv{\tilde{\varepsilon}}_{\vv{t}})$.
These continuous and phase-dependent terms are expanded into individual strain components, written as
\begin{align}\label{eq:strainB_voigtEX}
 (\vv{\varepsilon}_{\text{el}:\vv{n}}^{\alpha}+  \vv{\tilde{\varepsilon}}_{\vv{n}}^{\alpha}) & \equiv \left[ (\varepsilon_{\text{el}:{nn}}^{\alpha}+  \tilde{\varepsilon}_{nn}^{\alpha}), 2(\varepsilon_{\text{el}:{ns}}^{\alpha}+  \tilde{\varepsilon}_{ns}^{\alpha}), 2(\varepsilon_{\text{el}:{nt}}^{\alpha} +  \tilde{\varepsilon}_{nt}^{\alpha}) \right], \\ \nonumber
 (\vv{\varepsilon}_{\text{el}:\vv{t}}+  \vv{\tilde{\varepsilon}}_{\vv{t}}) & \equiv \left[ (\varepsilon_{\text{el}:{ss}}+  \tilde{\varepsilon}_{ss}), (\varepsilon_{\text{el}:{tt}}+  \tilde{\varepsilon}_{tt}), 2(\varepsilon_{\text{el}:{st}}+  \tilde{\varepsilon}_{st})  \right].
\end{align}
By employing the jump condition in Eqn.~\eqref{eq:force_balance1}, the stress components in the Voigt notation are additionally expressed as
\begin{align}\label{eq:stress_nt} 
 \vv{\sigma}_{\vv{B}}^{\alpha}=(\underbrace{\sigma_{nn}, \sigma_{ns}, \sigma_{nt}}_{:\equiv\vv{\sigma}_{\vv{n}}},\underbrace{ \sigma_{ss}^{\alpha}, \sigma_{tt}^{\alpha}, \sigma_{st}^{\alpha}}_{:\equiv\vv{\sigma}_{\vv{t}}^{\alpha}} )^{T},
\end{align}
where~$\vv{\sigma}_{\vv{n}}$ and~$\vv{\sigma}_{\vv{t}}^{\alpha}$ correspond to the continuous and phase-dependent stress components.

The elastic free energy density of phase~$\alpha$, which is based on the strain components transformed in the base~${\vv{B}}$, is written as
\begin{align}\label{eq:totalEL_free} 
 W_{\text{el}}^{\alpha}(\vv{\varepsilon}^{\alpha}_{\vv{B}})=\frac{1}{2}\left[ (\vv{\varepsilon}^{\alpha}_{\vv{B}}-\vv{\tilde{\varepsilon}}_{\vv{B}}^{\alpha})\cdot \C_{\vv{B}}^{\alpha}(\vv{\varepsilon}^{\alpha}_{\vv{B}}-\vv{\tilde{\varepsilon}}_{\vv{B}}^{\alpha}) \right],
\end{align}
where~$\C_{\vv{B}}^{\alpha}$ is the corresponding stiffness tensor.
The form of the stiffness tensor, adopted in Eqn.~\eqref{eq:totalEL_free}, is delineated in~\ref{AP:stiffness}.

For the present description of the energy density in Eqn.~\eqref{eq:totalEL_free}, the overall elastic free energy of the system is expressed as
\begin{align}\label{eq:elastic2}
 W_{\text{el}}(\vv{\varepsilon},\vphi)=\sum_{\alpha}W_{\text{el}}^{\alpha}(\vv{\varepsilon}^{\alpha}_{\vv{B}})\phia,
\end{align}
where the energy contributions of the individual phases are interpolated using the phase field, instead of a dedicated, monotonically varying functional.
The approach of employing the phase field as an interpolation function is established, and its accuracy in the current framework is elucidated elsewhere~\cite{amos2018chemo}.

The energy densities associated with the bulk phases principally contribute to the driving forces of the phase transformation.
In the present model, the elastic driving force is formulated as a pairwise interaction between two phases.
Therefore, the elastic contribution, governing the evolution between phase~$\alpha$ and~$\beta$, is written as
\begin{align}\label{eq:elastic_drivingforce1}
 \Delta W_{\text{el}}^{\alpha\beta}(\vv{\varepsilon}) =   \frac{\partial W_{\text{el}}(\vv{\varepsilon},\vphi)}{ \partial \phib} - \frac{\partial W_{\text{el}}(\vv{\varepsilon},\vphi)}{ \partial \phia}. 
\end{align}
The continuous variables which were identified earlier on the basis of the jump conditions are introduced to the model, by appropriately defining the elastic driving force $\Delta W_{\text{el}}^{\alpha\beta}(\vv{\varepsilon})$.
Consequently, the Legendre transform of the elastic free energy density is considered, which yields a corresponding elastic potential primarily dictated by the continuous variables.
The resulting elastic driving force, which directs the evolution of the phase field, is written as
\begin{align}\label{eq:elastic_drivingforce2}
 \Delta W_{\text{el}}^{\alpha\beta}(\vv{\varepsilon}_{t},\vv{\sigma}_{n}) & =  \frac{\partial P(\vv{\varepsilon}_{t},\vv{\sigma}_{n},\vphi)}{\partial \phib}   -
  \frac{\partial P(\vv{\varepsilon}_{t},\vv{\sigma}_{n},\vphi)}{\partial \phia}. 
\end{align}
The overall elastic potential $P(\vv{\varepsilon}_{t},\vv{\sigma}_{n},\vphi)$, dictating the driving force, reads
\begin{align}\label{eq:legendre_intp1}
 P(\vv{\varepsilon}_{t},\vv{\sigma}_{n},\vphi)& =    \Bigg\{  \left[ 
\begin{pmatrix} 
\vv{\sigma}_n \\ \vv{\varepsilon}_t 
\end{pmatrix} 
\cdot 
\bar{\TT}
\begin{pmatrix}
  \vv{\sigma}_n \\ \vv{\varepsilon}_t
\end{pmatrix} 
  \right]  \\ \nonumber
 & - \sum_{\alpha} \left[ 
\begin{pmatrix} 
  \vv{\sigma}_n \\ \vv{\varepsilon}_t 
\end{pmatrix} 
  \cdot 
\begin{pmatrix} 
  \vv{I} & \TT^\alpha_{nt}        \\
  \vv{O} & \TT^\alpha_{tt}   
\end{pmatrix}
\begin{pmatrix} 
  \tilde{\vv\varepsilon}^\alpha_n \\ \tilde{\vv\varepsilon}^\alpha_t 
\end{pmatrix} 
  \right]  
   + \frac{1}{2} \sum_{\alpha} \left( \tilde{\vv\varepsilon}^{\alpha}_t \cdot \TT^\alpha_{tt} \tilde{\vv\varepsilon}^{\alpha}_t\right) \Bigg\} \phia.
\end{align}
The elastic potential is extensively described by introducing a proportionality matrix~$\bar{\TT}$, which is separately discussed in~\ref{AP:stiffness}.

In the present framework, which involves the elastic potential, the normal and tangential stresses are calculated from the corresponding strain, through the proportionality matrix.
Therefore, the elastic and tangential components of the inelastic eigenstrains are ascertained to enable the formulation of the stresses.
The components of the inelastic strains are determined by
\begin{align}\label{eq:chi_n_t_poly}
  \tilde{\vv\chi}_n = \sum_\alpha \left(\tilde{\vv{\varepsilon}}^\alpha_n + \TT^\alpha_{nt}\tilde{\vv{\varepsilon}}^\alpha_t \right) \phia  \quad \text{and} \quad  \tilde{\vv\chi}_t = \sum_\alpha \TT^\alpha_{tt} \tilde{\vv{\varepsilon}}^\alpha_t \phia,
\end{align}
where~$\tilde{\vv\chi}_n$ and~$\tilde{\vv\chi}_t$ represent the normal and tangential component of the interpolated stress-free strains.
Using the predefined proportionality matrix, the stresses are calculated as follows from the components of the overall and inelastic strain:
\begin{align}\label{eq:sigma_B_multi}
\vv{\bar{\sigma}}_{B} = 
 \begin{pmatrix}
  - \bar{\TT}_{nn}^{-1}                & -\bar{\TT}_{nn}^{-1} \bar{\TT}_{nt} \\
  - \bar{\TT}_{tn} \bar{\TT}_{nn}^{-1} & \bar{\TT}_{tt} - \bar{\TT}_{tn} \bar{\TT}_{nn}^{-1} \bar{\TT}_{nt}
  \end{pmatrix} 
 \begin{pmatrix} 
 \vv{\varepsilon}_n \\ \vv{\varepsilon}_t 
 \end{pmatrix}
  + 
    \begin{pmatrix}
  \bar{\TT}_{nn}^{-1}                & \vv{O} \\
  \bar{\TT}_{tn} \bar{\TT}_{nn}^{-1} & - \vv{I}
  \end{pmatrix} 
 \begin{pmatrix} 
 \tilde{\vv{\chi}}_n \\ \tilde{\vv{\chi}}_t 
 \end{pmatrix}.
\end{align}
While the regular formulation can be adopted to calculate the stresses in the bulk phases, it is important to realize that the above Eqn.~\eqref{eq:sigma_B_multi} is necessary to estimate the stresses in regions where the phase field exhibits a spatial dependency.

In the current investigation, which is in accordance with the displacive theory~\cite{hehemann1970phase,bhadeshia1981rationalisation}, the growth of the bainite subunit is treated as a diffusionless transformation.
Since such a consideration obviates the need for a conventional description of a chemical free energy density, involving the evolution of concentration~\cite{amos2019phase,amos2018globularization}, a constant driving force is introduced in the bulk contribution~\cite{schoof2018multiphase}.
This constant chemical contribution is expressed as
\begin{align}\label{eq:chem_freeEnergy}
 \Delta W_{\text{ch}}^{\alpha\beta}(c,T, \vphi)=  \left[\frac{\partial W_{\text{ch}}(\vphi)}{\partial \phib}\right]_{c,T} - \left[\frac{\partial W_{\text{ch}}(\vphi)}{\partial \phia}\right]_{c,T} , 
\end{align}
where the driving force~$\Delta W_{\text{ch}}^{\alpha\beta}$ is the difference in the free energy densities of the phases with the constant carbon concentration~$c$ and the constant temperature~$T$.
Moreover, as in Eqn.~\eqref{eq:elastic2}, the overall chemical free energy is formulated of the interpolation of the individual phase contribution:
\begin{align}\label{eq:chem_sum}
 W_{\text{ch}}(c,T, \vphi)=\sum_{\alpha}W_{\text{ch}}^{\alpha}(c,T)\phia.
\end{align}
The quantitative difference in the free energy densities of the phases is involved by incorporating appropriate information from the TCFe8-CALPHAD database~\cite{andersson2002thermo}.

\subsubsection{Evolution equations}

Having appropriately defined the interface and the bulk contribution, the system is allowed to evolve towards a phenomenological decrease in the overall energy density~\cite{provatas2011phase}.
The phase-field evolution, which causes this progressive decrease in the energy density, formulated as a functional in Eqn.~\eqref{eq:functional}, is expressed as 
\begin{align}\label{eq:phase_field}
 \frac{\partial \phia}{\partial t} = -\frac{1}{\epsilon}\frac{1}{\tilde{N}} \sum_{\alpha<\beta}^{\tilde{N}} M_{\alpha\beta} \left[\frac{\delta \mathcal{F}(\vphi, \n \vphi,\vv{\varepsilon},c,T)}{\delta \phia} - \frac{\delta \mathcal{F}(\vphi, \n \vphi,\vv{\varepsilon},c,T)}{\delta \phib} \right],
\end{align}
where the mobility of the phase field is governed by~$M_{\alpha\beta}$.
For the present approach, this temporal evolution of the phase field reads as
\begin{align}\label{eq:phase_field2}
\frac{\partial \phia}{\partial t} = -\frac{1}{\epsilon}\frac{1}{\tilde{N}} \sum_{\alpha<\beta}^{\tilde{N}} M_{\alpha\beta}\left[ \frac{\delta \mathcal{F}_\text{intf}(\vphi, \n \vphi)}{\delta \phia} - \frac{\delta \mathcal{F}_\text{intf}(\vphi, \n \vphi)}{\delta \phib} - \frac{8 \sqrt{\phia \phib}}{\pi} \left( \Delta W^{\alpha \beta}_\text{chem} + \Delta W^{\alpha \beta}_\text{el} \right) \right],
\end{align}
where~$\tilde{N}$ is the number of active phases, as opposed to the total number~$N$~\cite{steinbach2009phase}.
Moreover, in Eqn.~\eqref{eq:phase_field2}, the variational derivative of the functional associated with the interface, $\mathcal{F}_\text{intf}(\vphi, \n \vphi)$, involves a derivation, pertaining to the phase field and its gradient, ($\partial \mathcal{F}_\text{intf}/\partial \phia-\n\cdot\partial \mathcal{F}_\text{intf}/\partial \n\phia$).
By introducing the elastic and chemical driving force, delineated in Eqns.~\eqref{eq:elastic_drivingforce2} and~\eqref{eq:chem_freeEnergy}, into Eqn.~\eqref{eq:phase_field2}, the phase field evolves without any excess contribution to the interface.

Since the phase transformations analyzed in this work are primarily governed by undercooling, the phase field is assumed to evolve in a mechanical equilibrium.
This mechanical equilibrium is imposed by
\begin{align}\label{eq:elastic}
\rho \frac{\partial^2 \vv{u}}{\partial t^2}=\n\cdot\vv{\sigma}=\vv{0},
\end{align}
where~$\rho$ is the mass density and~$(\n\cdot\vv{\sigma})_{i}$, the divergence in the stress, is determined by~$(\n\cdot\vv{\sigma})_{i}=\partial \sigma_{ij}/\partial x_j$.
Based on Eqn.~\eqref{eq:elastic}, the evolution of the dynamic variables, associated with the elastic contribution, is evaluated.

\subsection{Domain configuration}

In order to gain a fundamental understanding of the influence of the stress-free transformation on the autocatalytic growth of the subunits, all analyses are confined to two-dimensional setups of largely identical dimensions.
These two-dimensional simulation domains are discretized with equidistant voxel cells of the dimensions $\Delta x=\Delta y=0.65$~nm.
The entire domain consists of~$1199 \times 2400$ cells, thereby rendering a dimension of~$0.78 \times 1.56~\upmu$m.
Two different schemes are adopted to solve the evolution of the phase fields and the dynamic variables governing the elastic driving force.
While the phase-field evolution in Eqn.~\eqref{eq:phase_field2} is solved over the discretized domain, by the explicit forward-marching Euler scheme, the mechanical equilibrium in Eqn.~\eqref{eq:elastic} is treated implicitly in a staggered manner, using a finite element scheme.
Moreover, the entire domain is decomposed using the MPI~(Message Passing Interface) standard to reduce simulation time.

Periodic boundary conditions are assigned to the two-dimensional domain.
However, for solving the elastic variables, a plain strain condition is assumed in the third dimension.
Fixing the length scale parameter at~$\epsilon=3\Delta x$, a definite interface width of about seven cells is used in all simulations.
Because of the displacive nature of the subunit growth, and consequently, the high rate of the transformation, a dimensionless time step~$\Delta t=1$ is involved in the discussion.
The mobility governing the phase-field evolution is appropriately defined to enable a stable temporal evolution~\cite{schoof2018multiphase}.
All simulations in this work are performed with the multiphysics in-house software package \pace~(Parallel Algorithms for Crystal Evolution in 3D)~\cite{hotzer2018parallel}.

\subsection{Thermodynamic condition}

Following the seminal work on the bainite transformation, an analogous thermodynamic condition, involving an isothermal evolution in an Fe-C system with a carbon mole fraction of~$0.01968$, at~$573.15$~K, is considered for all numerical investigations~\cite{bhadeshia1981rationalisation}.
Due to the primary focus of explicating the influence of elasticity on the autocatalysis, the concentration evolution from the supersaturated subunits to the neighboring austenite matrix is overlooked in the present analysis. 
Since the growth of the subunits is significantly faster than the observable evolution of the concentration, such a treatment, excluding the carbon migration, is reasonable in the initial stages of the transformation.
Despite the lack of a dedicated consideration for carbon diffusion, the quantitative driving force pertaining to the given concentration and temperature, which dictates the growth of the bainite subunits, is derived from the TCFe8-CALPHAD database and is incorporated into the approach~\cite{schoof2018multiphase}.

\subsection{Subunit nuclei and cut-off area}

In all simulations, a stable elliptical nucleus of the dimension~$59.8 \times 5.85$~nm, which approximately yields an aspect ratio of ten, is considered as the precursor for a bainite subunit.
It is assumed that during the transformation, the bainite subunits follow the Bain orientation relationship as martensite~\cite{bhadeshia2019bainite}. 
Thus, the stress-free transformation strain
\begin{align}\label{eq:bain_strain}
\tilde{\varepsilon}^{\alpha}=
    \begin{pmatrix}
  \tilde{\varepsilon}^{\alpha}_{3}		& 0							&		0 \\
  0							& \tilde{\varepsilon}^{\alpha}_{1}	&		0 \\
  0							& 0							&		0	
  \end{pmatrix}
\end{align}
is introduced to provide the elastic driving force.
It is well established that the shear components predominantly dictate the eigenstrain associated with the bainite transformations~\cite{matsuzaki1994stress,swallow1996high}.
Therefore, the components of the eigenstrain in Eqn.~\eqref{eq:bain_strain} are quantified as~$\tilde{\varepsilon}^{\alpha}_{3}=0.1$ and~$\tilde{\varepsilon}^{\alpha}_{1}=-0.1$, involving a prefactor to include any plastic accommodation.
As indicated earlier, a single Bain variant is exclusively considered, since bainite subunits exhibit an identical orientation relation~\cite{bhadeshia2019bainite}.
For the formulation of the elastic driving force, bainite and austenite are treated as isotropic, linear elastic phases, with a Young's modulus of~$E=210$~GPa and a Poisson's ratio of~$v=0.3$.
The phases are rotated in such a way that the preferred growing direction lies in a direction of a coordinate axis~\cite{amos2018chemo}.
The interfacial energy between the phases is fixed at~$\gamma=0.2$~Jm$^{-2}$~\cite{luzginova2008bainite}.

During the bainite transformation, the growth of a subunit is impeded by the accumulation of a plastic strain at the tip~\cite{bhadeshia1980mechanism}.
Because of not including the plastic strain into the present model, the restricted growth is achieved by introducing a \lq cut-off area\rq \thinspace.
When the bainite subunit reaches this cut-off area, its progressive growth is prevented.
The cut-off area is appropriately defined to ensure the minimum influence of the boundary condition on a given subunit or its neighbor.
Correspondingly, the growth of a bainite subunit is impeded when its area fraction in the domain is about~$0.006$.
In this work, the identical cut-off area is adopted for all simulations. 

\section{Result and discussion}

\subsection{Elastic interaction}
\begin{figure}[!ht]
    \centering
      \includegraphics[]{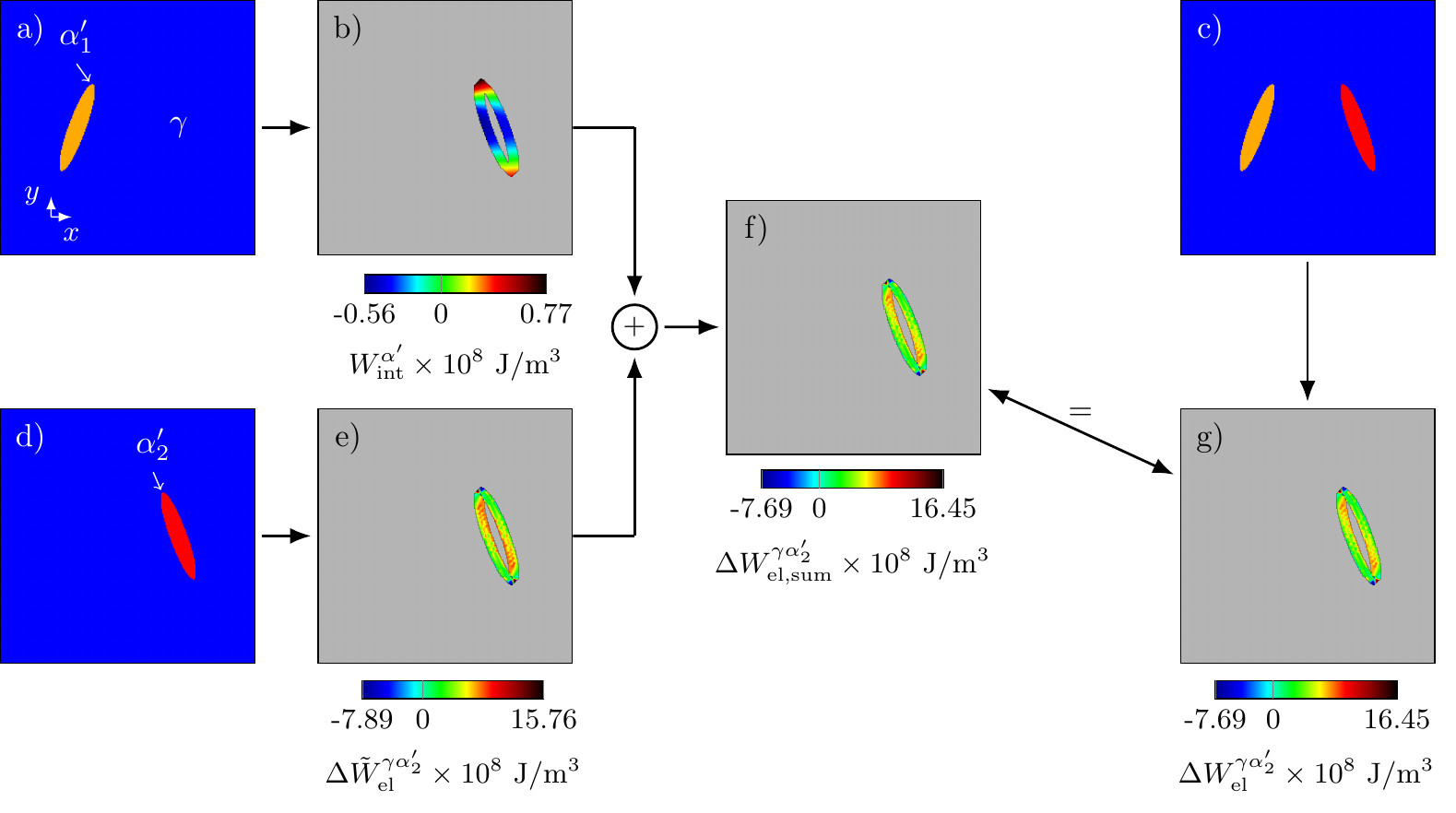}
    \caption{Representative analysis to explain the recovery of the elastic interaction.
             a) and d) Primary structures of bainite subunits, placed at different spatial locations, in the absence of any pre-existing subunit.  
             b) \textit{Induced} elastic energy, according to Eq.~\eqref{eq:induced}, in a region to be occupied by the second subunit. 
             c) Combined microstructure of a) and d).
             e) Elastic driving force, according to Eq.~\eqref{eq:elastic_drivingforce2}, \textit{inherent} to the growth of a subunit, in the absence of any pre-existing structures. 
             f) Sum of b) and e).
             g) Elastic driving force, according to Eq~\eqref{eq:elastic_drivingforce2}, resulting from the interaction between the induced and inherent elastic energy.
             The elastic energy in b) and e)-g) is depicted in the interface region of austenite and the second subunit~($\alpha'_2$).
             In case of the setting c), the elastic driving force of the second subunit can be analyzed by either using the microstructure c) or calculating the sum of the energies b) and e), based on settings a) and d).
    \label{fig:fig2validation}}
\end{figure}

The stress-free transformation strain contributes to the elastic driving force, which is \textit{inherently} associated with the subunits~\cite{bhadeshia1981rationalisation}.
During the growth of the primary subunit, this inherent driving force is entirely dictated by the eigenstrain.
However, the secondary subunits grow in an elastically strained environment, which is due to the presence of the primary subunit.
Therefore, the driving force governing the evolution of the secondary subunits is the result of the interaction between the inherent driving force, based on the eigenstrain, and the spatially inhomogeneous stress, \textit{induced} by the primary subunit~\cite{zhou2010effect,shi2013variant}.
To understand the influence of elasticity on the growth of the subunits, this interaction should be recovered sufficiently.
In order to verify the ability of the present approach, which is the provision of the elastic interaction between the subunits, a representative case is analyzed, as shown in Fig.~\ref{fig:fig2validation}.
For the sake of generality, the eigenstrain defined by the components $\varepsilon_1 = -0.08$ and $\varepsilon_3 = 0.12$ is adopted for this preliminary investigation.

%%%%%%%%%%%%%%%%%%%%% figure start  %%%%%%%%%%%%%%%%%%%%%
%%%%%%%%%%%%%%%%%%%%% figure end  %%%%%%%%%%%%%%%%%%%%%

As shown in Fig.~\ref{fig:fig2validation}a, a stable bainite subunit is placed in the domain.
Due to the corresponding stress-free transformation strains, a nonuniform, elastic environment is introduced in the matrix, which is calculated by solving the mechanical equilibrium~\eqref{eq:elastic}.
Instead of analyzing the inhomogeneous stress, established across the entire domain, the elastic interaction energy, calculated as the product of the local stress and the eigenstrain,
\begin{align}\label{eq:induced}
W_\text{int}^{\alpha'}(\vv x)=-\vv{\sigma}(\vv{x})\tilde{\vv{\varepsilon}}^{\alpha'},
\end{align}   
is estimated for the region \textit{to be occupied} by the subsequent subunit.
The distribution of the elastic energy, calculated through Eqn.~\eqref{eq:induced}, in the region pertaining to the subsequent subunit, is illustrated in Fig.~\ref{fig:fig2validation}b.

As shown in Fig.~\ref{fig:fig2validation}d, a second subunit is placed in a different position of the domain, which is completely devoid of any pre-existing subunits.

The inherent driving force of the second independent subunit, $\Delta \tilde{W}_\text{el}^{\gamma\alpha'_2}$, which is based on the eigenstrain, is determined through Eqn.~\eqref{eq:elastic_drivingforce2} and is plotted in Fig.~\ref{fig:fig2validation}e.
In the presence of the primary subunit, as shown in Fig.~\ref{fig:fig2validation}c, the elastic driving force established across the second subunit changes to $\Delta {W}_\text{el}^{\gamma\alpha'_2}$, which is illustrated in
Fig.~\ref{fig:fig2validation}g, due to its interaction with the induced stress field of the primary subunit.
Correspondingly, the driving force associated with the secondary subunit can be expressed as
\begin{align}\label{eq:interaction1}
\Delta W_\text{el}^{\gamma\alpha'_2}(\vv x) = \Delta \tilde{W}_\text{el}^{\gamma\alpha'_2}(\vv x) +  W_{\text{int}}^{\alpha'}(\vv x).
\end{align}
It is evident from Figs.~\ref{fig:fig2validation}g and~\ref{fig:fig2validation}f, which shows the sum calculated from the results depicted in Figs.~\ref{fig:fig2validation}b and~\ref{fig:fig2validation}e, that the present approach elegantly recovers the elastic interaction in the above Eqn.~\eqref{eq:interaction1}.
Therefore, this technique is employed to investigate the role of elasticity in the autocatalytic growth of the secondary subunits.

\subsection{Favorable interaction contour}
\begin{figure}[!ht]
    \centering
      \includegraphics[]{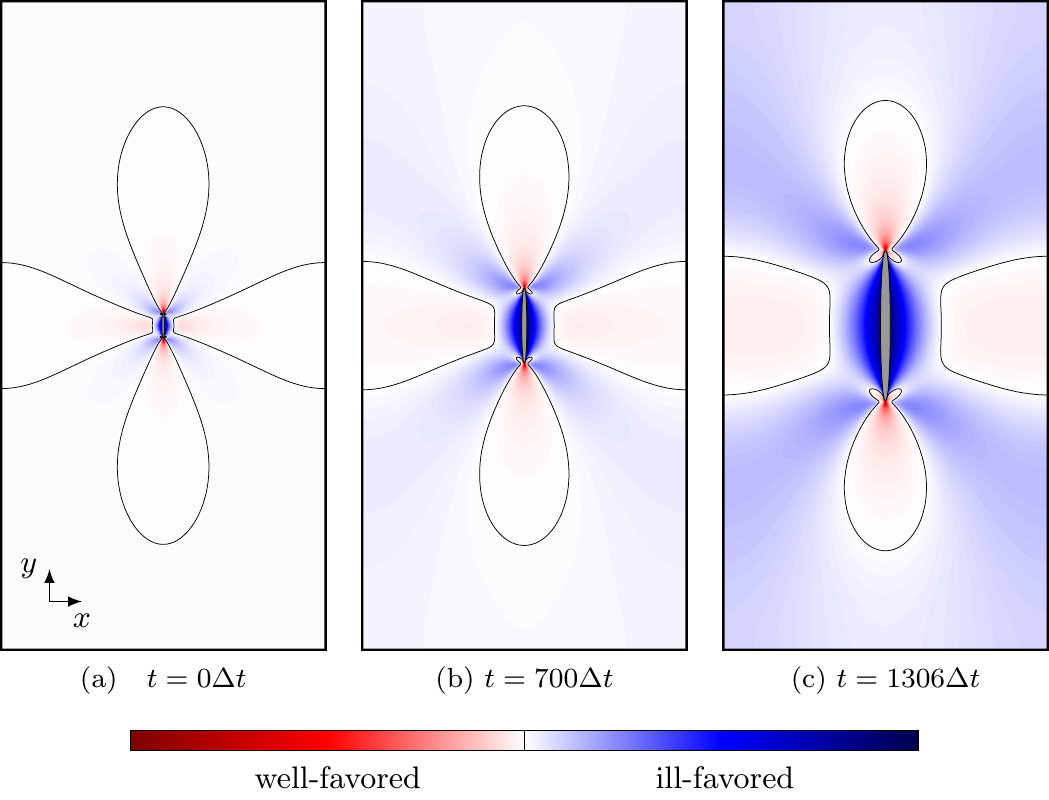}
    \caption{Temporal evolution of the \textit{favorability contour,} during the growth of the primary subunit.
    \label{fig:fig3criterion}}
\end{figure}

The nonuniform, elastic interaction energy, induced by the eigenstrain of the growing primary subunit, is illustrated in Fig.~\ref{fig:fig3criterion}.
%%%%%%%%%%%%%%%%%%%%% figure start  %%%%%%%%%%%%%%%%%%%%%
%%%%%%%%%%%%%%%%%%%%% figure end  %%%%%%%%%%%%%%%%%%%%%
In a conventional analytical treatment, this elastic energy is quantified and incorporated with the thermodynamic criterion for the subsequent growth of the subunits~\cite{bhadeshia1982bainite,rees1992bainite}.
Such a treatment of elastic energy, induced by the stress-free transformation strains, overlooks its favorable contribution, which enhances the free energy density, dictating the growth of the bainite subunit.
This favorable interaction can be formulated by considering the interaction between the inherent and the induced elastic contribution.

As elucidated in the previous section, the elastic driving force substantially governs the evolution of the bainite subunit.
This elastic driving force,~$\Delta W_\text{el}^{\gamma\alpha'_1}$, is dictated by the eigenstrain associated with the subunit.
In addition to the driving force, the growth of the primary subunit introduces a spatially inhomogeneous stress into the austenite matrix.
The elastic energy, induced by the inhomogeneous stress, can be quantified through Eqn.~\eqref{eq:induced}. 
Under this condition, \textit{i.e}, in the presence of the primary subunit, the growth of the second subunit is elastically favored when
\begin{align}\label{eq:el_favor}
 \Delta W_\text{el}^{\gamma\alpha'_2}(\tilde{\varepsilon}^{\alpha}) +  W_{\text{int}}^{\alpha'}(\vv{x}) < \Delta W_\text{el}^{\gamma\alpha'_1}(\tilde{\varepsilon}^{\alpha}),
\end{align}
where~$\Delta W_\text{el}^{\gamma\alpha'_2}$ is the elastic driving force, involved with the second subunit.
Since the driving forces of both the primary and the secondary subunits are dictated by the identical eigenstrain and periodic boundary conditions are applied, $\Delta W_\text{el}^{\gamma\alpha'_1}(\tilde{\varepsilon}^{\alpha}) = \Delta W_\text{el}^{\gamma\alpha'_2}(\tilde{\varepsilon}^{\alpha})$ holds.
From Eqn.~\eqref{eq:el_favor}, the criterion governing the elastically preferred growth can therefore be expressed as
\begin{align}\label{eq:el_favor1}
 W_{\text{int}}^{\alpha'}(\vv x)<0.
\end{align} 
Based on the above criterion, a \textit{favorability contour}, characterized by~$W_{\text{int}}^{\alpha'}(\vv x)=0$, is plotted in Fig.~\ref{fig:fig3criterion}.
Within the region with~$W_{\text{int}}^{\alpha'}(\vv x)<0$, the growth of the subunit is energetically more favored, due to the elastic interaction.
In other words, the interaction between the induced energy of the primary subunit and the inherent energy of the secondary one aides the growth of the second subunit within the favorability contour.
It is vital to realize that this contour pertains to the growth of the identical variants, and not to the complementary ones, as observed during the martensite transformation.

Fig.~\ref{fig:fig3criterion} shows that the contour which demarcates the favorable interaction correspondingly evolves with the growth of the primary subunit.
With the expansion of the subunit, it can be seen, in particular, that the region along the broad faces, wherein the growth of a nucleus with an identical variant is preferred, visibly moves away from it.
In contrast, the favorability contour, which is confined to the tip of the subunit, continues to exist, despite the temporal evolution. 
Moreover, an ear-like morphology is adopted by the favorable region at the tip of the bainite subunit, when it reaches the cut-off area.
The observed configuration of the elastically preferred contour is consistent with the previous analytical work, wherein the shape-dependent elastic interaction was predicted on the basis of the seminal work on the elastic stability of an inclusion~\cite{eshelby1957determination,eshelby1959elastic,olson1976stress}.

\subsection{Gradient in elastic interaction}
\begin{figure}[!ht]
    \centering
      \includegraphics[]{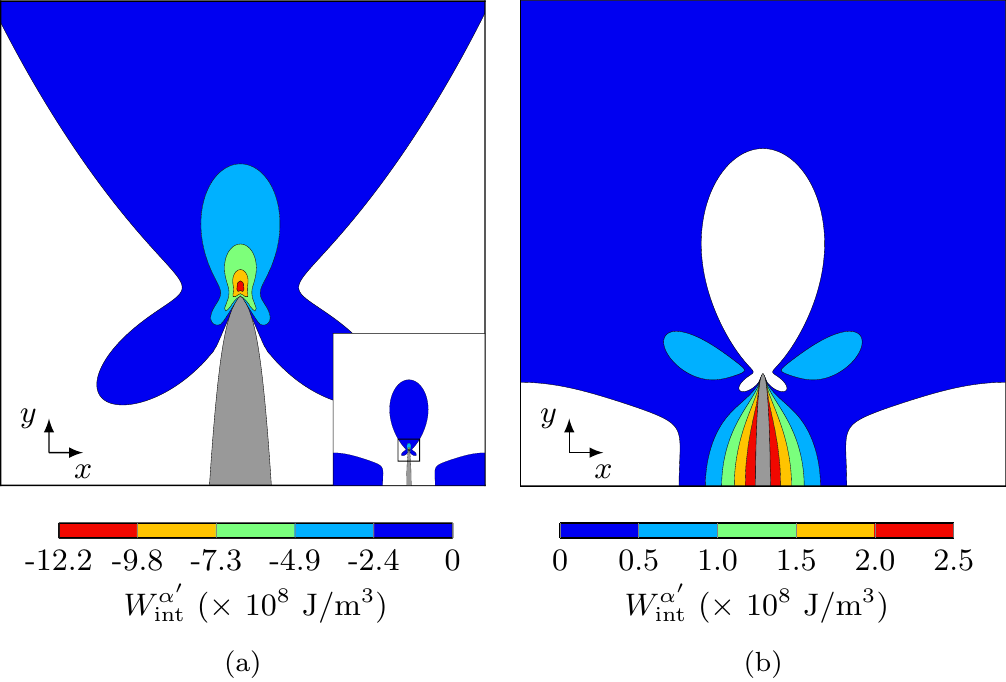}
    \caption{Spatial dependence of the interaction energy a) at the tip and b) along the broad faces of the primary subunit.
    The microstructure corresponds to the result depicted in Fig.~\ref{fig:fig3criterion}c.
    \label{fig:fig5gradient_df}}
\end{figure}

In order to quantitatively identify the spatially dependent interaction between the primary and secondary subunit, the induced elastic energy, surrounding the primary subunit, is quantified.
Fig.~\ref{fig:fig5gradient_df}a illustrates the induced elastic energy, which is calculated as a product of local stress and eigenstrain~(Eqn.~\eqref{eq:induced}), within the favored region around the tip of the subunit.
Accordingly, the induced energy that contributes to the growth kinetics of the secondary subunit noticeably varies within the elastically preferred region.

Fig.~\ref{fig:fig5gradient_df}a suggests that the positions closer to the tip are highly favorable for the autocatalysis of the subunit, when compared to the other location.
%%%%%%%%%%%%%%%%%%%%% figure start  %%%%%%%%%%%%%%%%%%%%%
%%%%%%%%%%%%%%%%%%%%% figure end  %%%%%%%%%%%%%%%%%%%%%
With an increase in the distance from the tip, the negative interaction energy, which renders an elastically favored autocatalytic growth, progressively diminishes.
In Fig.~\ref{fig:fig5gradient_df}a, the calculation and the depiction of the induced elastic energy absolutely agree with the spatially varying transformation kinetics, illustrated in Fig.~\ref{fig:fig4tip_growth}.

Analogous to the calculation of the induced energy within the favorability contour at the tip, the elastic interaction energy in the ill-favored region, which is adjacent to the broad faces of the full-grown primary subunit, is determined and shown in Fig.~\ref{fig:fig5gradient_df}b.
This illustration reemphasizes that the induced elastic energy does not enhance the growth of the secondary subunits, beyond the favorability contour.
In other words, the elastically preferred autocatalytic growth is confined to the contours, and in the remaining regions, the elastic interaction negatively influences the evolution of the secondary structures.
In these ill-favored regions, the growth of the secondary subunit is only plausible when the driving force is high enough to overcome the elastic constraint.

As illustrated in Fig.~\ref{fig:fig5gradient_df}b, the negative effect of the resultant elastic energy is spatially dependent, which is similar to the effect of the positive interaction energy.
In a full-grown subunit, for instance, the regions immediately close to the broad faces are ill-favored for the growth of the secondary subunits.
However, in positions away from the broad faces, the degree of the positive interaction energy decreases.
Beyond a critical distance, the induced elastic energy offers a negative interaction energy, which favors the growth of the secondary subunit, thereby indicating the transition to the elastically preferred region. 

By combining the elastic driving force, altered by the primary subunit, with the existing chemical contribution, as shown in Figs.~\ref{fig:fig3criterion} and~\ref{fig:fig5gradient_df}, an effective driving force can be calculated, which dictates the growth of the secondary subunits.
The effective driving force can indeed be employed to refine the calculation of the activation energy~($Q_{\text{A}}$), which, as indicated by Eqn.~\eqref{eq:frame3a}, is also involved in determining the nucleation rates of secondary subunits~\cite{van2008modeling,gaude2006new,ravi2017bainite}.
However, such an estimation would notably be restricted, owing to the principal consideration of the present analysis.
In other words, according to the displacive theory of the bainite transformation, the growth of the supersaturated subunit is followed by the rejection of carbon to the surrounding austenite matrix.
This partitioning of carbon significantly changes the effective driving force, by influencing the chemical contribution.
Therefore, estimating the effective driving force in the current framework, which reasonably overlooks the carbon diffusion to ascertain the elastically governed autocatalysis, would be limited to a considerable extent and and cannot be directly adopted in the quantitative calculation of the nucleation rate.
Nevertheless, in the upcoming works, which encompass carbon partitioning, the combined effect of the spatially inhomogeneous stress and concentration fields on the activation energy will be analyzed, and the resulting change in the nucleation rate will be discussed.

\subsection{Favored growth kinetics}

\begin{figure}[!ht]
    \centering
      \includegraphics[]{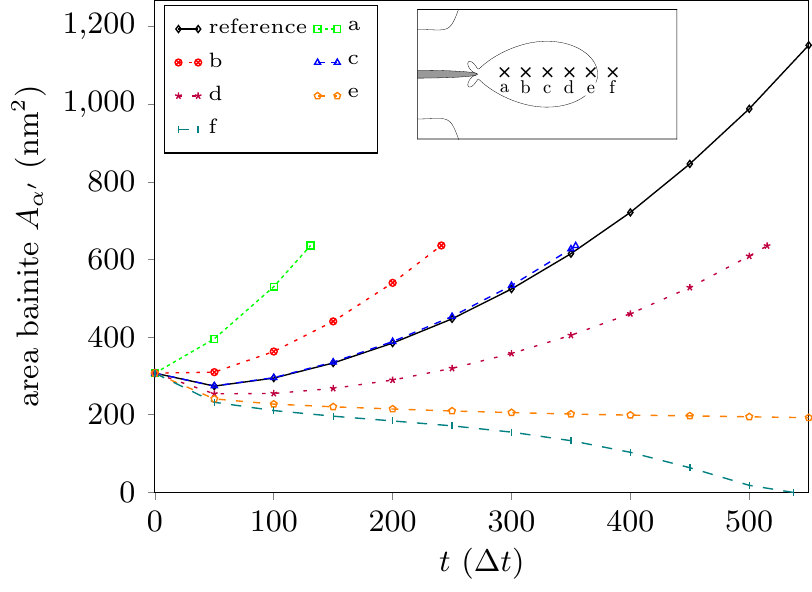}
    \caption{Growth rate, plotted as the temporal change in the phase fraction (area) of the second subunit, with varying spatial positions.
    \label{fig:fig4tip_growth}}
\end{figure}

While the thermodynamic criterion in Eqn.~\eqref{eq:el_favor1} convincingly renders a criterion for the elastically preferred growth of the subunits, an understanding of the degree of favorability cannot be extracted from this relation.
In other words, for a given chemical driving force, the influence of the favorable interaction on the growth of the autocatalytic subunit cannot be ascertained from the thermodynamic relation in Eqn.~\eqref{eq:el_favor1}.
Therefore, a second subunit is allowed to evolve at a different location within the favorable region, which is confined to the tip of the primary structure, as shown in the subset of Fig.~\ref{fig:fig4tip_growth}.
%%%%%%%%%%%%%%%%%%%%% figure start  %%%%%%%%%%%%%%%%%%%%%
%%%%%%%%%%%%%%%%%%%%% figure end  %%%%%%%%%%%%%%%%%%%%%
In order to clarify the role of the elastic interaction, the chemical contribution, which dictates the growth rate of the second subunit, is considered to be less than that of the primary subunit 
\begin{align}\label{eq:chem_condition}
\Delta W_{\text{ch}}^{\gamma\alpha'_{2}}=0.91\Delta W_{\text{ch}}^{\gamma\alpha'_{1}},
\end{align}
where~$\Delta W_{\text{ch}}^{\gamma\alpha'_{2}}$ and~$\Delta W_{\text{ch}}^{\gamma\alpha'_{1}}$ are the chemical driving forces, respectively governing the transformation rate of the secondary and the primary subunit.
In the absence of a favorable elastic interaction, the condition in Eqn.~\eqref{eq:chem_condition} leads to the shrinking of the subunit.
Therefore, any deviation from this reasonably expected evolution results from the local elastic interaction, which enables the autocatalytic growth of the bainitic subunit. 

The growth rate of the secondary subunit, positioned at different locations at the tip of the primary structure, is plotted in Fig.~\ref{fig:fig4tip_growth}, by monitoring the change in the bainite area.
Evidently, the transformation kinetics varies with the position of the second subunit.
This difference in the growth rate indicates a spatially dependent elastic interaction. 
Fig.~\ref{fig:fig4tip_growth} furthermore shows that despite the lower chemical driving force, the subunits located at the position~$a$, $b$ and~$c$ either exhibit a higher growth rate than the primary subunits or an equal one. 
The enhanced growth of the secondary subunits is principally due to the favorable interaction of the elastic energy.

As illustrated in Fig.~\ref{fig:fig4tip_growth}, the kinetics of the growth of the secondary subunit decreases correspondingly, as its position moves towards the end of the favorability contour.
Ultimately, at the location~$f$, which is outside the elastically preferred region, the stable nucleus of the subunit begins to shrink and eventually disappear.

\subsection{Elastically preferred autocatalytic growth}
Having demarcated the elastically well- and ill-favored region for the autocatalytic growth, and recognizing its spatial dependency in Fig.~\ref{fig:fig5gradient_df}, the effect of the elastic interaction, in the actual evolution of the secondary structures, is analyzed by monitoring the evolution of stable nuclei at different locations around a full-grown primary bainite subunit.
Because of the four-fold symmetry of the domain, and the corresponding energy distribution, the present investigation is restricted to a quadrant.

\begin{figure}[!ht]
    \centering
      \includegraphics[width=1.0\textwidth]{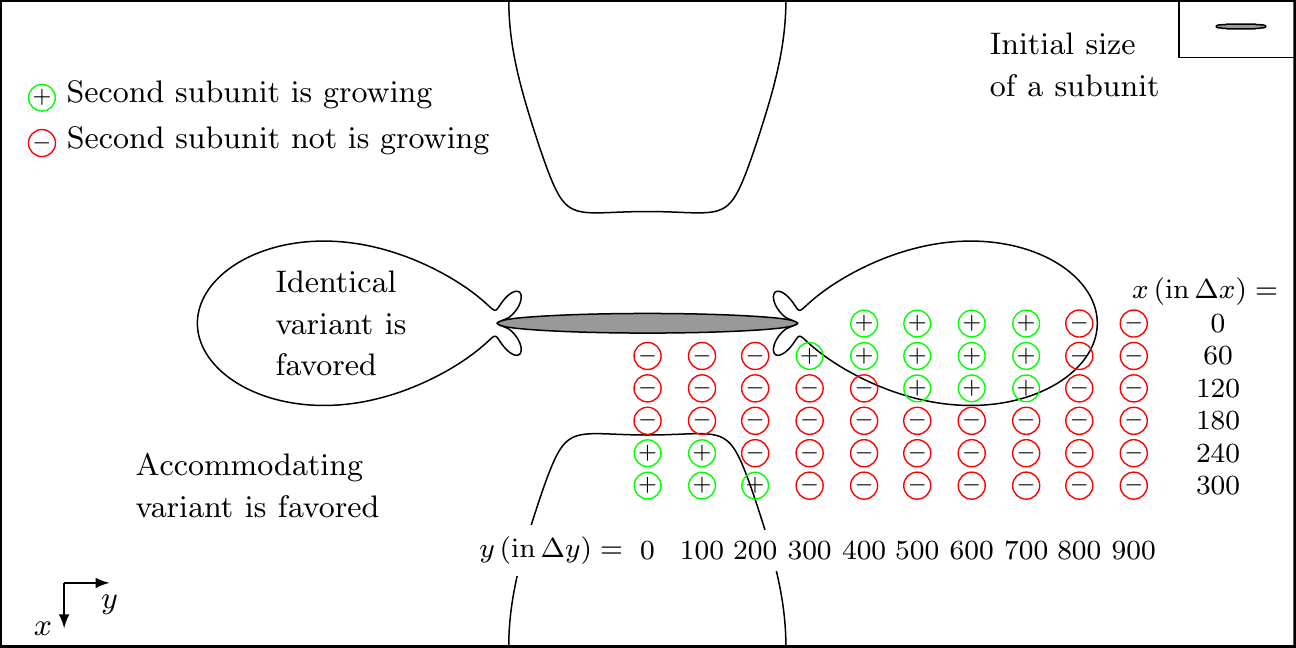}
    \caption{Evolution of stable nuclei in a quadrant of the symmetric domain. 
             The growth of the subunit is distinguished from its decay by appropriate symbols, which are included in the illustration.
    \label{fig:fig6allover}}
\end{figure}

Different spatial positions, considered for analyzing the autocatalytic elastic interaction, are shown in Fig.~\ref{fig:fig6allover}.
%%%%%%%%%%%%%%%%%%%%% figure start  %%%%%%%%%%%%%%%%%%%%%
%%%%%%%%%%%%%%%%%%%%% figure end  %%%%%%%%%%%%%%%%%%%%%
Moreover, based on the nature of the evolution, a distinction is made between different locations, in this illustration.
The symbol \lq +\rq \thinspace, enclosed by a green circle, indicates the growth of the nucleus, while the symbol \lq -\rq \thinspace, encircled in red, depicts a shrinkage and a decay of the stable precursor. 

In complete agreement with the favorability contour plotted in Eqn.~\eqref{eq:el_favor1}, solely based on the thermodynamic condition, the nuclei within the contour exhibit growth, while the ones outside shrink and ultimately disappear.
As indicated in Eqn.~\eqref{eq:chem_condition}, it is important to note that the chemical driving force for the growth of the secondary subunits is lower than the driving force for the primary structure.
This low driving force, which is introduced to explicate the role of the elastic interaction, is responsible for the slight deviations that are observed at the rims of the favorability contours, confined to the tip.
The analysis illustrated in Fig.~\ref{fig:fig6allover} suggests that, as opposed to the conventional description of the autocatalysis as the growth of the subunits from the bainite-austenite interface, for a fully grown primary structure, the elastically preferred growth is restricted to the tip and specific regions visibly away from the broad faces.

\subsection{Evolution of the contour with secondary subunit growth}

\subsubsection{At the tip}
Despite the lack of a comprehensive understanding of the influence of the elastic interaction on the growth of subunits, it has been realized that the tip of the primary structure is a preferred site for the growth of the subsequent precursors~\cite{tszeng2000autocatalysis,gaude2006new}.
Since several nucleation and growth events, which are influenced by the elastic interaction, follow the evolution of the second subunit, the temporal change in the morphology of the favorability contour is determined, which accompanies the growth of the secondary structure.

\begin{figure}[!ht]
    \centering
      \includegraphics[]{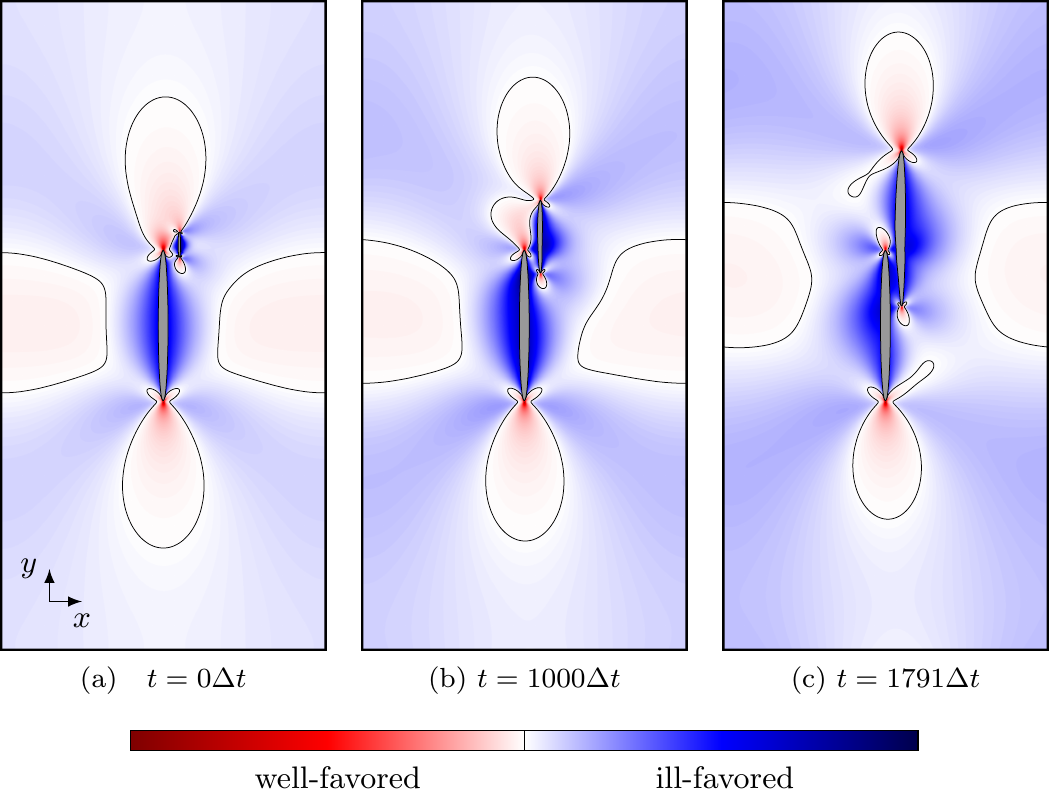}
    \caption{Change in the morphology of the favorability contour, with the introduction of a second subunit, and its temporal evolution, with the growth of the nucleus.
    \label{fig:fig7tip_criterion}}
\end{figure}

The change in the spatial distribution of the regions, which offers a negative elastic interaction energy, is illustrated in Fig.~\ref{fig:fig7tip_criterion}, with the evolution of the second precursor, at the tip of the full-grown primary subunit.
Despite the abundance of possible spots which, as shown in Fig.~\ref{fig:fig6allover}, render an elastically favored growth of the secondary subunit, a position displaced from the tip is chosen to reflect the established understanding on the arrangements of subunits in a bainite sheave~\cite{bhadeshia2019bainite} and to corroborate with the outcomes of the existing numerical model~\cite{arif2013phase}.
%%%%%%%%%%%%%%%%%%%%% figure start  %%%%%%%%%%%%%%%%%%%%%
%%%%%%%%%%%%%%%%%%%%% figure end  %%%%%%%%%%%%%%%%%%%%%
In order to capsulate this evolution, the nucleus is premeditatedly placed within the favorable contour at the tip.
Therefore, the negative elastic interaction energy enables the growth of the secondary subunit, despite the relatively low chemical driving force (Eqn.~\eqref{eq:chem_condition}).

As shown in Fig.~\ref{fig:fig7tip_criterion}, an independent favorability contour, associated with the secondary subunit, is induced in the initial stages of the evolution.
This negative interaction region, exclusively pertaining to the second subunit, is visible at its lower tip.
On the other end, it can be observed that at the upper tip of the secondary unit, for example, the pre-existing elastic energy, which is established by the primary structure, interacts with the one induced by the growing secondary subunit. 

As the second subunit grows, the elastically preferred region, associated with the primary structure, decreases.
Proportionately, the area enclosed by the favorability contour of the secondary unit expands with its evolution, as shown in Fig.~\ref{fig:fig7tip_criterion} at~$t=1000\Delta t$.
Despite this expansion of the elastically preferred region at one tip of the second subunit, the corresponding configuration of the contour at the opposite tip, adjacent to the primary structure, remains unchanged.
The elastically favored sections on either side of the primary subunits slightly shift upwards.
Moreover, the elastically preferred region, adjacent to the broad face of the second subunit, disturbs the symmetry of the configuration by protruding inwards, as shown in Fig.~\ref{fig:fig7tip_criterion}, at~$t=1000\Delta t$.

When the evolving subunit reaches the cut-off area, as shown in Fig.~\ref{fig:fig7tip_criterion}, at~$t=1791\Delta t$, the symmetry in the distribution of the elastic energy is reestablished. 
A unique configuration is respectively adopted by the favorability contour at the upper and lower tip of the secondary and primary subunits.
Furthermore, the area of the elastically preferred region, which is parallel to the broad faces of the subunits, is increased by their inward protrusion. 
When compared to the initial configuration, an upward shift is noticeable in this region.
In Fig.~\ref{fig:fig7tip_criterion}, the evolution of the subsequent subunits is dictated by the contours enclosing the negative interactions at~$t=1791\Delta t$.

\subsubsection{Along the broad faces}
\begin{figure}[!ht]
    \centering
      \includegraphics[]{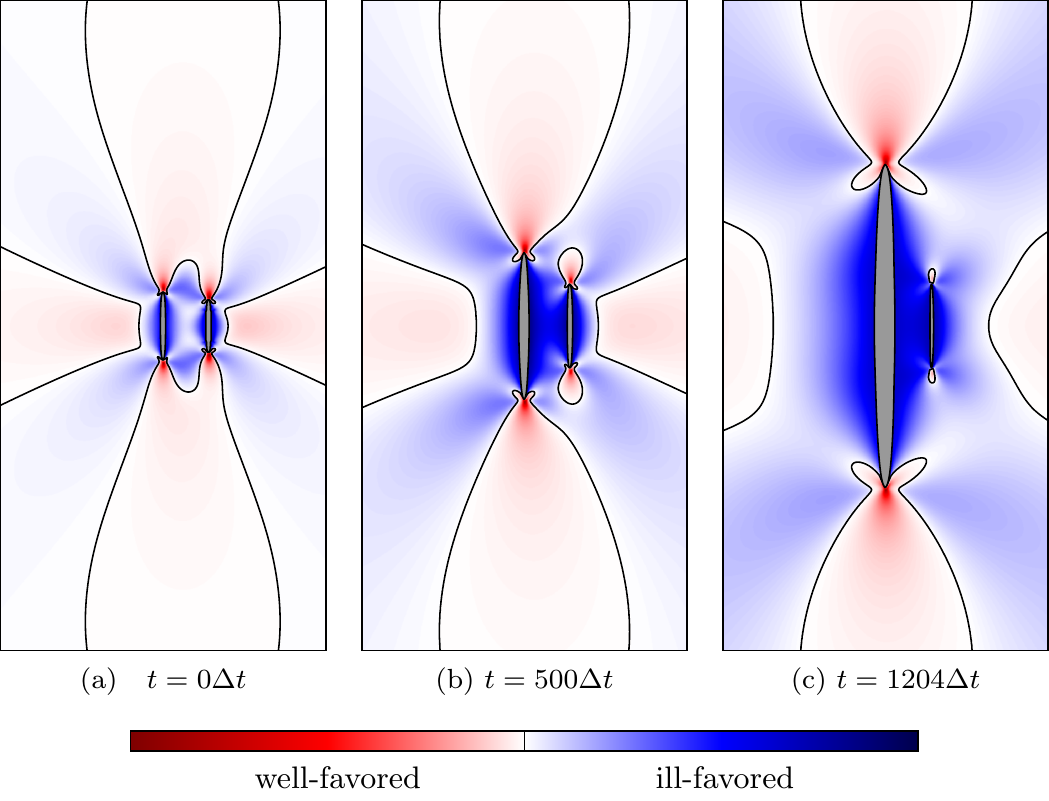}
    \caption{Restricted growth and ultimate shrinking of the secondary nucleus, during the parallel evolution of the subunits. 
             The driving force for the growth of the second subunit is less than that of the first, in accordance with Eqn.~\eqref{eq:chem_condition}.
             To make the second subunit more visible, it is pointed out that the depicted snapshots do not represent the entire domain.
    \label{fig:fig8parallel_shrink}}
\end{figure}

In addition to the tip, another location which is commonly perceived as the favored spot for the autocatalytic growth of bainite subunits is along the broad faces of the primary structure~\cite{bhadeshia1980mechanism}.
During martensite transformation, complementing variants grow in a accommodating manner, by sharing a common interface and, consequently, compromising the individual morphology~\cite{bhattacharya1992self}.
In contrast, the bainite subunits appear to be distinct, while adhering to their characteristic shape.
In Fig.~\ref{fig:fig3criterion}, a previous investigation on the evolution of the favorability contour, surrounding the primary subunit, indicates that when the nucleus reach the cut-off area, the region favoring the parallel growth of the subunits is significantly separate.
Yet, in the early stages of the evolution, the growth of the secondary subunit, adjacent to the broad face of the primary bainite, is elastically favored, as shown in Fig.~\ref{fig:fig3criterion}, at~$t=0$.
To analyze the parallel evolution of the subunits, the secondary precursor therefore is introduced to the elastically preferred region, adjacent to the broad face of the primary nucleus, after its marginal growth.
The evolution of the subunits, wherein the secondary nucleus occupy a position within the favorability contour, adjacent to the broad faces of the primary subunit, is shown in Fig.~\ref{fig:fig8parallel_shrink}.
%%%%%%%%%%%%%%%%%%%%% figure start  %%%%%%%%%%%%%%%%%%%%%
%%%%%%%%%%%%%%%%%%%%% figure end  %%%%%%%%%%%%%%%%%%%%%
The driving force dictating the growth of the secondary precursor is much smaller than the driving force of the primary precursor, satisfying the condition in Eqn.~\eqref{eq:chem_condition}.
In Fig.~\ref{fig:fig6allover} and Fig.~\ref{fig:fig7tip_criterion}, it has previously been shown that the negative elastic interaction enables the growth of the second subunit, despite the low chemical driving force.
However, in the parallel configuration, shown in Fig.~\ref{fig:fig8parallel_shrink}, the primary nucleus evolves at a much faster rate than the secondary nucleus.
Despite the difference in the transformation rate, both subunits grow in the initial stages of the transformation.
Evidently, as illustrated in Fig.~\ref{fig:fig8parallel_shrink}, at~$t=1204\Delta t$, the evolution of the secondary nucleus is reversed, when the primary subunit becomes sufficiently large. 
Moreover, as the primary nucleus continue to expand, the secondary nucleus begin to shrink and ultimately disappear.
Despite its presence in the elastically preferred region, the disappearance of the secondary subunit can be attributed to the initial difference in the chemical driving force.

The initial difference in the chemical driving force, as shown in Fig.~\ref{fig:fig8parallel_shrink}, transfers a higher transformation rate to the primary subunit.
The favorability contour, which evolves with the subunits, consequently moves away from the position of the secondary nucleus, which is due to the dominant growth of the primary subunit.
Ultimately, the disparity in the growth kinetics locally transforms the growth of the secondary nucleus into an ill-favored one.
As shown in Fig.~\ref{fig:fig8parallel_shrink}, the favorability contour therefore fails to evolve parallelly in the secondary subunit, adjacent to the broad face of the primary structure.

\begin{figure}[!ht]
    \centering
      \includegraphics[]{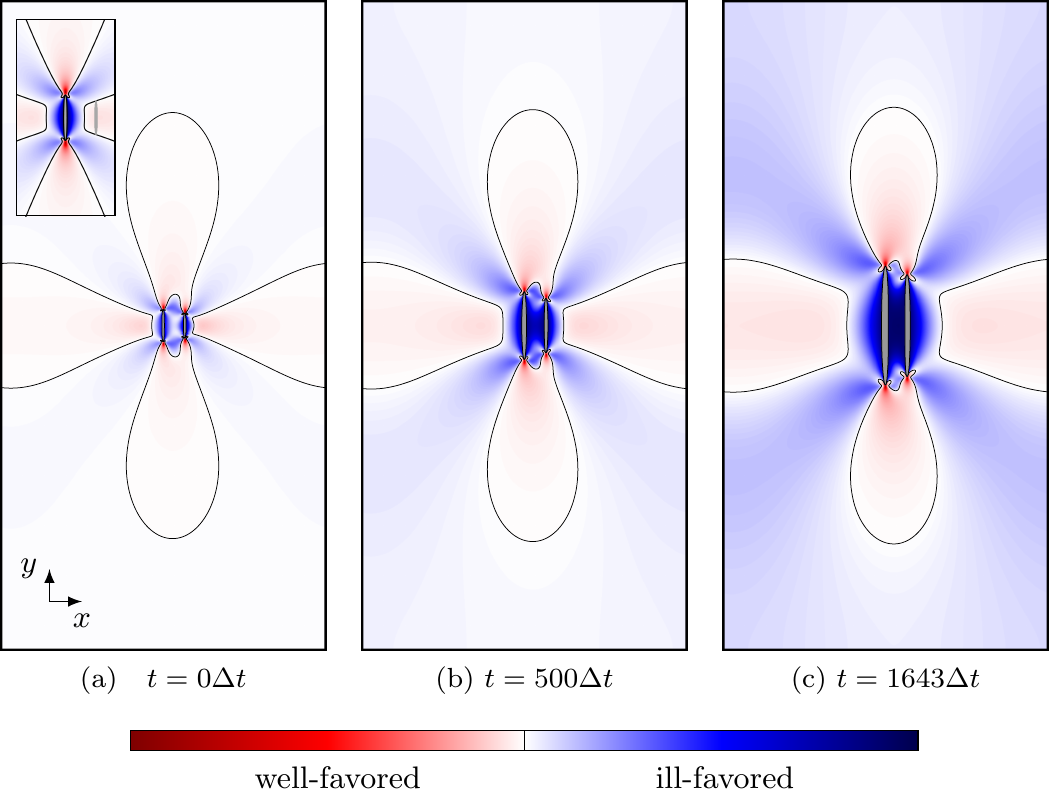}
    \caption{Parallel growth of the primary and secondary subunit, which is achieved by relaxing the condition in Eqn.~\eqref{eq:chem_condition} and incorporating an equal chemical driving force.
    \label{fig:fig8parallel_growth}}
\end{figure}

In order to overcome the disparity in the transformation kinetics, equal chemical driving forces are assigned to both the primary and secondary subunit, exclusively for this investigation.
By adopting the similar initial configuration, as in Fig.~\ref{fig:fig8parallel_shrink}, but using the equal chemical contribution, the evolution is illustrated in Fig.~\ref{fig:fig8parallel_growth}.
%%%%%%%%%%%%%%%%%%%%% figure start  %%%%%%%%%%%%%%%%%%%%%
%%%%%%%%%%%%%%%%%%%%% figure end  %%%%%%%%%%%%%%%%%%%%%
As opposed to the outcomes of the previous consideration, the secondary subunit, with an equal driving force, continues to grow along the broad face of the primary structure. 
Moreover, the favorability contours at the tips of these subunits interact and provide a unified preferred region.
Fig.~\ref{fig:fig8parallel_growth} additionally shows that the elastically favored regions are much closer to the broad faces of these primary and secondary subunits, when compared to the isolated primary structure in Fig.~\ref{fig:fig3criterion}.

\section{Conclusion}

In the present work, the effect of the stress-free transformation strain on the autocatalytic growth of the subunits is extensively analyzed by adopting an elastic phase-field model which recovers the configurational force balance.
In the following, the insights gained by the current investigations are enumerated:
\begin{itemize}
\item The influence of the eigenstrain, which is often viewed as a criterion to be overcome by the driving force~\cite{bhadeshia1982bainite,rees1992bainite}, is realized to be spatially dependent and to favor the subunit growth at specific locations, despite the low chemical contribution.
Based on the elastic interactions, these regions are distinguished through a favorability contour, within which the growth of the subunits is elastically more preferred.
In addition to demarcating the elastically favored regions, the evolution of the subunits within and across these sections is examined by monitoring the growth or decay of several nuclei at different spatial positions.
\item Even within the favored region, the effect of the elasticity on the growth kinetics varies because of the inhomogeneous distribution of the stresses.
\item The commonly conceived growth of the subunits at the tip is attributed to the preferred interaction between the elastic energy, induced by the primary subunit, and is inherent to the secondary one. 
However, the favorable interaction at the bainite-austenite interface is only confined to a small fraction of the tip.
In other words, energetically preferred growth of the secondary subunits at the tip is more restricted than is usually postulated.
\item For the parallel growth of the subunits along the broad face of the primary structure, the secondary subunit should nucleate at the early stages of the transformation, so that its location is elastically preferred.
Additionally, since the favorability contour moves away from the broad faces, with the growth of the primary subunit, the growth rate of the secondary nucleus should be comparable to that of the primary one.
\item With the growth of the primary and secondary subunits, the elastically favored regions characteristically evolve and enables a preferred growth of subsequent nuclei.
The contours become partially continuous with its introduction to the Bain strain.
\end{itemize}
The above conclusions indicate that the role of elasticity in the autocatalysis of the bainite transformation is rather convoluted and cannot be encompassed comprehensively by the stored energy criterion.
However, based on the present investigation, the formulation for ascertaining the autocatalytic nucleation rate can be refined, thereby enhancing the analytical treatment of the bainite transformation kinetics.

In the current study, the role of plasticity, which is the prevention of the growth of the subunit, is replaced by an imposed cut-off area.
Therefore, attempts are made to investigate the bainite transformation, using an elastoplastic model, in order to explain the role of plastic accommodation.
Moreover, the concentration evolution and its influence on the chemical driving force are not considered in the present analysis.
This aspect of the bainite transformation will be addressed in the future, along with the effect of a volumetric component of the eigenstrain.

\section*{Acknowledgments}

P.G. Kubendran Amos thanks for the financial support of the German Research Foundation (DFG), under the project AN 1245/1. 
The other authors acknowledge the financial support of the Ministry of Science, Research and Arts of Baden-Wuerttemberg, under the grant 33-7533.-30-10/25/54.
The authors are also grateful for the editorial support by Leon Geisen.
Parts of this work were performed on the computational resource ForHLR II, funded by the Ministry of Science, Research and Arts of Baden-Wuerttemberg and the DFG.

\section*{Conflicts of interest}
The authors declare no conflicts of interest.

\appendix
\section{Stiffness tensor and proportionality matrix}\label{AP:stiffness}
Following Refs.~\cite{schneider2015phase,schneider2018small}, the stiffness tensor in the coordinate system~${\vv{B}}$, which is associated with the elastic free energy density formulation, is expressed as 
\begin{align}\label{eq:C_nt}
\C^{\alpha}_B &=
\begin{pmatrix}
\begin{array}{ccc|ccc}
\mathcal{C}_{nnnn}^{\alpha} & \mathcal{C}_{nnnt}^{\alpha} & \mathcal{C}_{nnns}^{\alpha} & \mathcal{C}_{nntt}^{\alpha} & \mathcal{C}_{nnss}^{\alpha} & \mathcal{C}_{nnts}^{\alpha}\\
\mathcal{C}_{ntnn}^{\alpha} & \mathcal{C}_{ntnt}^{\alpha} & \mathcal{C}_{ntns}^{\alpha} & \mathcal{C}_{nttt}^{\alpha} & \mathcal{C}_{ntss}^{\alpha} & \mathcal{C}_{ntts}^{\alpha}\\
\mathcal{C}_{nsnn}^{\alpha} & \mathcal{C}_{nsnt}^{\alpha} & \mathcal{C}_{nsns}^{\alpha} & \mathcal{C}_{nstt}^{\alpha} & \mathcal{C}_{nsss}^{\alpha} & \mathcal{C}_{nsts}^{\alpha}\\ \hline
\mathcal{C}_{ttnn}^{\alpha} & \mathcal{C}_{ttnt}^{\alpha} & \mathcal{C}_{ttns}^{\alpha} & \mathcal{C}_{tttt}^{\alpha} & \mathcal{C}_{ttss}^{\alpha} & \mathcal{C}_{ttts}^{\alpha}\\
\mathcal{C}_{ssnn}^{\alpha} & \mathcal{C}_{ssnt}^{\alpha} & \mathcal{C}_{ssns}^{\alpha} & \mathcal{C}_{sstt}^{\alpha} & \mathcal{C}_{ssss}^{\alpha} & \mathcal{C}_{ssts}^{\alpha}\\
\mathcal{C}_{tsnn}^{\alpha} & \mathcal{C}_{tsnt}^{\alpha} & \mathcal{C}_{tsns}^{\alpha} & \mathcal{C}_{tstt}^{\alpha} & \mathcal{C}_{tsss}^{\alpha} & \mathcal{C}_{tsts}^{\alpha}
\end{array}
\end{pmatrix}.
\end{align}
For the purposes of numerical operation, this stiffness tensor is split into four blocks:
\begin{align}\label{eq:C_nt2}
\C^{\alpha}_B &=
\begin{pmatrix}
\C_{nn}^{\alpha} & \C_{nt}^{\alpha}\\
\C_{tn}^{\alpha} & \C_{tt}^{\alpha}
\end{pmatrix}.
\end{align}
The proportionality matrix, involved in the formulation of the elastic potential ($P(\vv{\varepsilon}_{t},\vv{\sigma}_{n},\vphi)$), is similarly written as
\begin{align}\label{eq:T_nt}
 \bar{\TT}=
\begin{pmatrix}
\bar{\TT}_{nn} & \bar{\TT}_{nt}\\
\bar{\TT}_{tn} & \bar{\TT}_{tt}
\end{pmatrix}.
\end{align}
Each entity of the proportionality matrix corresponds to an interpolated block and reads
\begin{align}
\bar{\TT}_{nn} &:= \sum_\alpha \TT^\alpha_{nn} \phia := - \sum_\alpha {(\C^\alpha_{nn})}^{-1} \phia \\ 
\bar{\TT}_{nt} &:= \sum_\alpha \TT^\alpha_{nt} \phia :=  \sum_\alpha {(\C^\alpha_{nn})}^{-1} \C^\alpha_{nt} \phia  \\
\bar{\TT}_{tt} &:= \sum_\alpha \TT^\alpha_{tt} \phia := 
\sum_\alpha \left(\C^\alpha_{tt} - \C^\alpha_{tn} {(\C^\alpha_{nn})}^{-1} \C^\alpha_{nt} \right) \phia.
\end{align}

\section*{References}

\bibliographystyle{elsarticle-num}
\bibliography{library.bib}
\end{document}